\documentclass[prb,twocolumn,floatfix]{revtex4}

\newcommand{\ket}[1]{|#1\rangle}
\newcommand{\bra}[1]{\langle#1|}

\newcommand{\pj}[1]{\ket{#1}\bra{#1}}

\newcommand{\ep}{\varepsilon}
\newcommand{\nn}{\nonumber}
\usepackage{graphicx}
\usepackage{amssymb,bm}

\usepackage{amsmath}

\begin{document}
\title{Junctions of one-dimensional quantum wires -- correlation effects in
  transport}
\author{X.\ Barnab\'e-Th\'eriault}\thanks{Xavier B.-T. passed away in a tragic 
traffic accident on August 15, 2004.}
\affiliation{Institut f\"ur Theoretische  Physik, Universit\"at G\"ottingen, 
Friedrich-Hund-Platz 1, D-37077 G\"ottingen, Germany}
\author{A.\ Sedeki}
\affiliation{Institut f\"ur Theoretische  Physik, Universit\"at G\"ottingen, 
Friedrich-Hund-Platz 1, D-37077 G\"ottingen, Germany}
\author{V.\ Meden}
\affiliation{Institut f\"ur Theoretische  Physik, Universit\"at G\"ottingen, 
Friedrich-Hund-Platz 1, D-37077 G\"ottingen, Germany}
\author{K.\ Sch\"onhammer}
\affiliation{Institut f\"ur Theoretische  Physik, Universit\"at G\"ottingen, 
Friedrich-Hund-Platz 1, D-37077 G\"ottingen, Germany}

\begin{abstract} 
We investigate transport of spinless fermions through a 
single site dot junction of $M$ one-dimensional quantum wires.
The semi-infinite wires are described by a tight-binding model.
Each wire consists of two parts: the non-interacting leads and a region of
finite extent in which the fermions interact via a nearest-neighbor
interaction.
The functional renormalization group method is used to determine 
the flow of the linear conductance as a function of a low-energy cutoff 
for a wide range of parameters. Several fixed points are identified 
and their stability is analyzed. We determine the scaling exponents 
governing the low-energy physics close to the fixed points. Some of
our results can already be derived using the non-self-consistent
Hartree-Fock approximation. 
\end{abstract}

\maketitle     

\section{Introduction}
\label{Intro}

In one spatial dimension correlation effects strongly influence the
low-energy physics of many-fermion systems. Such systems cannot
be described as Fermi liquids, but are classified as
Tomonaga-Luttinger liquids (TLLs), which are characterized by a
vanishing quasi-particle weight and power-law scaling of correlation
functions.\cite{KS} For spin rotationally invariant interactions and
spinless fermions, on which we focus here, the exponents of the 
different correlation functions can be expressed in terms of a 
single number, the TLL parameter $K$. It depends on the parameters 
of the chosen model, in particular the strength of the two-particle 
interaction. For vanishing interaction $K=1$, while $0 < K < 1$ for 
repulsive interaction and $K>1$ in the attractive case. As indicated 
by the singular behavior of the density response function at momentum 
transfer $2 k_F$,\cite{LutherPeschel,Mattis} with $k_F$ the Fermi momentum, 
a TLL reacts quite differently to an inhomogeneity than a
Fermi liquid. The physics of inhomogeneous TLLs can  
conveniently be studied investigating transport properties.

The simplest junction is a single impurity in 
an infinite TLL wire. The transport through such a system has
intensively been studied in the past. Using the 
renormalization group (RG) language the single impurity problem 
can be characterized by two fixed points (FPs).\cite{KaneFisher} 
One is the ``perfect chain'' FP at which the
impurity effectively vanishes and the conductance takes its maximal 
value. For TLL wires that are ``smoothly'' coupled to non-interacting 
leads, a situation we consider here, the latter is given by
$G=e^2/h$.\cite{locallut} The
correction to the FP conductance asymptotically scales as 
$s^{2(K-1)}$, with $s$ being the largest (but still asymptotically
small) energy scale (e.g.~temperature, bias voltage, 
external infrared cutoff). 
For $0 < K <1$ the exponent is
negative and the FP is unstable, while it is stable for $K>1$. 
The other FP is the  ``decoupled chain'' FP at which $G=0$. The 
correction scales as $s^{2 \alpha_B}$, with $\alpha_B=1/K-1$. 
The FP is stable for repulsive interactions and
unstable in the attractive case. 
The exponent $\alpha_B$ characterizes the power-law behavior of the
local one-particle spectral weight of a TLL with an open boundary 
close to the boundary.\cite{KaneFisher,FG} 
The flow from one to the other FP is described by a $K$-dependent 
one-parameter scaling function.
The scaling behavior of the conductance has been demonstrated
for a simplified effective low-energy
model\cite{KaneFisher,Furusaki0,Moon,MatveevGlazman,Fendley} as well
as for a microscopic lattice model.\cite{VM4,Enss04}     
       
Recently single-walled carbon nanotubes were used to experimentally
realize junctions of several quasi one-dimensional quantum
wires.\cite{Fuhrer,Terrones} They might form the basis of future 
nano-electronic devices. Taking into account the fermion interaction, models
for different types of junctions and networks of TLLs have been 
investigated theoretically using a variety of 
methods.\cite{Nayak,Safi,Lal,Chen,Claudio,Das,Doucot,Xavier} 
These studies left open several interesting questions.
Already the low-energy physics of the three wire Y-junction is 
much richer than that of the single impurity problem. 

We here study the transport through a single site 
dot junction of $M=2,3,\ldots$ semi-infinite wires, each described 
by a microscopic lattice model, at temperature $T=0$.    
To obtain the conductance between the $M$ legs we mainly use an
approximate technique that is based on the
functional renormalization group (fRG)
method.\cite{Wetterich,Morris,Manfred} It has earlier been
successfully applied to describe the transport in a TLL with a single
impurity\cite{VM4,Enss04} and a double barrier,\cite{VMresotun,Enss04}
the latter allowing for resonant tunneling. The approximations lead to
reliable results for not too strong interactions with TLL parameter
$1/2 \leq K \leq 3/2$. In particular, for a single impurity the
power-law scaling of the conductance discussed above is reproduced 
with exponents that agree with the exact ones to leading order in 
the interaction. For the $M$-leg
junction we investigate the RG flow for a wide range of parameters and 
identify the FPs. We numerically determine the exponents of the 
power-law corrections to the FP conductances that govern the
low-energy physics close to the FPs. They depend on the 
interaction and the number of wires $M$. Most of these exponents have 
not been determined before. As in our approximation terms of second 
order in the interaction are only partly included the exponents 
can only  expected to be correct to leading order in the
interaction. In a short publication we have 
earlier verified that for a specific type of triangular three wire 
junction (not discussed in the present publication), for which an 
exact result is available,\cite{Claudio} we indeed reproduce the 
scaling exponent to leading order.\cite{Xavier} 

For a specific set of junction parameters the fRG study is  
supplemented by results for the conductance 
obtained using the non-self-consistent Hartree-Fock
approximation (HFA) and Fermi's golden rule like arguments. The HFA 
allows us to analytically calculate one of the scaling exponents. It 
has earlier been shown that this approximation leads to meaningful
results for the power-law scaling of the one-particle spectral weight 
in a TLL with an open boundary.\cite{OBCVM} 
      
The paper is organized as follows. In Sect.~\ref{model} we introduce
our model of the $M$-wire junction. The fRG based approximation scheme
is discussed in Sect.~\ref{method}. Using single-particle scattering
theory in this section we also derive equations relating the conductance
to matrix elements of an auxiliary Green function and the dot Green  
function. They can be used 
to reduce the numerical effort for solving the fRG flow equations and
to gain a deeper understanding of our findings for the conductance. 
In Sect.~\ref{HF} we apply the HFA to determine scaling exponents for a
certain class of symmetric junctions. Our fRG results for the FPs, the scaling
exponents of the corrections to the FP conductances, and the general
RG flow are presented in Sect.~\ref{fRGresults}. We conclude with a
summary and an outlook in Sect.~\ref{summary}. 

\section{The model}
\label{model} 

\begin{figure}[tbh]
\begin{center}
\includegraphics[width=0.35\textwidth,clip]{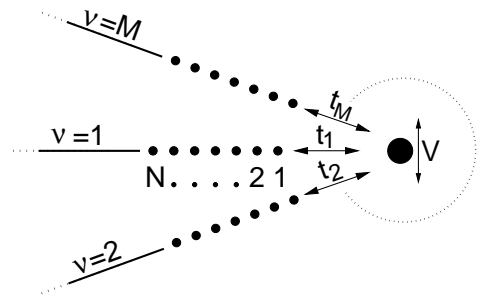}
\end{center}
\caption[]{A single site dot junction of $M$ quantum wires. Across the
  bonds of the lattice sites $j=1,\ldots,N$ (small filled circles) 
  the fermions 
  interact via a nearest-neighbor interaction, while they are
  non-interacting in the leads with $j>N$ (solid line). 
  The hopping amplitude between the first site of wire
  $\nu=1,\ldots,M$ 
  and the dot site is $t_\nu$. The on-site energy on the dot site (large filled
  circle) is $V$.\label{skiz}}
\end{figure}

Each of the $M$ quantum wires that meet at that single
site dot junction is described by the lattice model of spinless 
fermions with nearest-neighbor hopping. The semi-infinite wires can be 
divided in two sections: the lead with lattice sites $j>N$ in which 
the fermions are assumed to be non-interacting and the interacting wire with
nearest-neighbor interaction across the bonds of the sites $j \in [1,N]$.    
Fig.~\ref{skiz} shows a sketch of our system. We here focus on the
half-filled band case.
The results are generic also for other fillings. 

The Hamiltonian reads
\begin{equation}
  \label{fullH}
  H = H_{\rm kin}+H_{\rm int}+ H_{\rm junc}\; .
\end{equation}
The kinetic energy is modeled by
\begin{equation}
  H_{\rm kin}=-t \,\sum_{\nu=1}^{M}\sum_{j=1}^{\infty} \big( \,
  c^{\dag}_{j+1,\nu} c_{j,\nu}^{\phantom\dag} + 
c^{\dag}_{j,\nu} \, c_{j+1,\nu}^{\phantom\dag} \, \big)
\end{equation}
where we used standard second-quantized notation with $c^{\dag}_{j,\nu}$
and $c_{j,\nu}^{\phantom\dag}$ being creation and annihilation
operators on site $j$ of wire $\nu$, respectively. 
From now on we set $t=1$, i.e.\ measure energies in units of $t$.

As the part of the Hamiltonian containing the interaction we take
\begin{equation}
  H_{\rm int}=\sum_{\nu=1}^{M}\sum_{j=1}^{N-1} U_{j,j+1}
\left[ n_{j,\nu} - 1/2\right]   \left[ n_{j+1,\nu} - 1/2 \right]\; , 
  \label{wireH}
\end{equation}
with the local density $n_{j,\nu} = c^{\dag}_{j,\nu}\,
c_{j,\nu}^{\phantom\dag}$. The interaction $U_{j,j+1}$ is assumed to
be independent of the wire index and acts
only between the bonds of the sites $1$ to $N$, that define the
interacting wire. Within this region it is allowed to depend on the position. 
By subtracting the average filling $1/2$ from the density  $n_{j,\nu}$  
we prevent a depletion of the interacting part of the wire. 
The chemical potential corresponding to half-filling is $\mu=0$. 
To avoid
any fermion backscattering at the contact between the lead and the 
interacting wire, $U_{j,j+1}$ is turned on smoothly\cite{locallut}
starting at zero across the bond $(N,N+1)$ and approaching its bulk
value $U$ at bond $(N-j_s,N-j_s+1)$.\cite{VM4,VM5,Enss04} More explicitly
we use 
\begin{eqnarray}
\label{shape}
  U_{j,j+1} = \frac{U}{2} \left\{
  1-\frac{\arctan\left[s \;\pi (2[j-N]+j_s)/j_s \right]}
  {\arctan[s\,\pi]}\right\}
\end{eqnarray} 
for $j=N-j_s,...,N$ and $U_{j,j+1} = U$ for $1\leq
j<N-j_s$. The larger $N$ the smoother the interaction has to be switched
on. We here consider interacting wires of up to $N=10^5$ sites for which 
$j_s=32$ and $s=2$ turned out to be sufficient. 
For these parameters the backscattering at the lead-interacting wire
contact is less than $10^{-4} \%$ and can thus be neglected.
The results do not dependent on the detailed shape of the envelope 
function as long as it is sufficiently smooth. 

The model corresponding to the Hamiltonian 
$H_{\rm kin} + H_{\rm int}$ with interaction $U$ across 
all bonds (not only the ones within $[1,N]$) and $M=1$ shows TLL behavior 
for $|U| < 2$ with a TLL parameter (for half-filling)\cite{Haldane} 
\begin{eqnarray}
\label{K}
 K = \left[\frac{2}{\pi}  
 \arccos \left(-\frac{U}{2} \right) \right]^{-1} \; .
\end{eqnarray} 
To leading order in the interaction it is given by 
\begin{eqnarray}
\label{Kpert}
K= 1 - \frac{U}{\pi} + {\mathcal O}(U^2) \;  ,
\end{eqnarray} 
an expression we repeatedly refer to further down.

The junction we model by 
\begin{equation}
\label{dotH}
 H_{\rm junc} = -\sum_{\nu=1}^{M} t_{\nu} 
\big( \,c^{\dag}_{1,\nu}
 d^{\phantom\dag} + d^{\dag}\, c_{1,\nu}^{\phantom\dag} \, \big)+ 
V \,d^{\dag}d^{\phantom\dag}\; ,
\end{equation}
with $d^{\dag}$ and $d^{\phantom\dag}$ being creation and annihilation
operators on the dot site, respectively. 
It is parameterized by the hopping amplitudes $t_\nu \geq 0$ connecting the 
wire $\nu$ to the dot and the on-site energy $V \geq 0$ on the dot.
For $M=2$ the junction is equivalent to a local impurity 
in an infinite wire. Applying the fRG 
for this case we recover the 
results for the conductance obtained 
earlier (see below).\cite{VM4,Enss04}   

Note that in our Hamilltonian the fermions on the dot site do not 
interact with the fermions on the first lattice sites of the wires. 
Including such additional interactions does not lead to any
changes of the FP structure and scaling exponents
investigated here, as we have verified 
explicitly. We exclude such terms from our model as otherwise we later
would have to introduce renormalized junction parameters which would 
lead to an unnecessary proliferation of symbols.\cite{Enss04}

\section{The method}
\label{method}
At $T=0$ all inelastic processes are frozen out and 
the linear conductance $G_{\nu,\nu'}$ from wire $\nu$ to wire $\nu'$ 
can exactly be expressed in terms of a real space matrix element of 
the one-particle Green function ${\mathcal G}(\varepsilon+i0)$ evaluated at  
$\varepsilon=\mu=0$,\cite{Oguri,Enss04} 
\begin{eqnarray}
\label{conduct}
 \frac{h}{e^2} \;  G_{\nu,\nu'} = \left|t_{\nu,\nu'} \right|^2 = 
4|\langle N,\nu|{\mathcal G}(0+i0)|N,\nu'\rangle|^2 \; .   
\end{eqnarray}
Here $|N,\nu\rangle $ denotes the Wannier state centered on 
site $N$ of wire $\nu$ and 
$\left|t_{\nu,\nu'} \right|^2$ is the effective transmission from wire
$\nu$ to $\nu'$, with $\nu \neq \nu'$ and $\nu,\nu' \in [1,M]$. 
Note that ${\mathcal G}(\varepsilon+i0)$ must be calculated 
in the presence of the non-interacting leads, the junction,
and the interaction. 

\subsection{The functional renormalization group}

To obtain an approximation for the Green function
we use the fRG. A detailed account of the method was given in
Refs.\ \onlinecite{Enss04} and \onlinecite{Sabine}. We here only
present the approximate flow
equations (which are then integrated numerically), 
describe the most important steps to derive them, and give details 
specific to the present junction geometry.

An infrared cutoff $\Lambda$ is introduced by replacing the
non-interacting imaginary frequency propagator ${\mathcal G}_0$ of the 
system by the  $\Lambda$ dependent propagator 
 \begin{equation}
\label{cutoffproc}
{\mathcal G}_0^{\Lambda}(i \omega) =  \Theta(|\omega|-\Lambda) \, 
{\mathcal G}_0(i \omega) \; .
\end{equation}  
The cutoff runs from $\Lambda=\infty$ down to $\Lambda=0$, at which 
${\mathcal G}_0^{\Lambda=0}(i \omega) = {\mathcal G}_0(i \omega)$ is 
reached and the cutoff-free problem is recovered.
Using the generating functional for one-particle irreducible vertex
functions, with ${\mathcal G}_0^{\Lambda}$ as the non-interacting
propagator,  
an infinite hierarchy of coupled flow equations for the self-energy, 
the effective two-particle interaction, and higher order vertex
functions is derived.\cite{Wetterich,Morris,Manfred} It is truncated 
by neglecting the three-particle vertex, which is a valid 
approximation as long as the two-particle vertex does not 
become too large. The two-particle vertex projected on the Fermi 
points is parameterized by an effective nearest-neighbor interaction  
$U^\Lambda$. The flow equation for the latter is obtained by
considering a single infinite chain with interaction across all bonds
and neglecting self-energy corrections.\cite{Sabine} 
It can be integrated and at half-filling $U^{\Lambda}$ is given by
\begin{equation}
\label{flowU}
 U^{\Lambda} = \frac{U}
 {1 + 
 \left(\Lambda - \frac{2 + \Lambda^2}{\sqrt{4 + \Lambda^2}} \right) \, 
 U/(2\pi)} \; .
\end{equation}
The $\Lambda$-dependent two-particle vertex is then approximated 
by a frequency independent nearest-neighbor interaction of strength 
$U^{\Lambda}$. 
In the case where the interaction depends on position,
as an additional approximation we apply Eq.\ (\ref{flowU}) locally 
for each bond. 
As a consequence of the assumed frequency independence of the effective
interaction also the self-energy does not depend on 
$\omega$. In the exact solution an $\omega$-dependence is
generated to order $U^2$ (bulk TLL behavior). 
This exemplifies that in our approximation
for the self-energy terms of order $U^2$ are only partly included. 

With these
approximations the self-energy is diagonal in the wire index $\nu$ and
tridiagonal in the lattice site index $j$. 
In a next step the non-interacting leads are projected
out.\cite{Enss04} This results in an additional  
diagonal and Matsubara frequency $\omega$-dependent one-particle potential 
\begin{eqnarray}
\label{leadpotdef}
 \left<j,\nu\right|  V_{\rm lead}(i\omega)   \left| j',\nu'\right>  & = &
 \frac{i\omega}{2} \left( 1 - 
 \sqrt{1 + \frac{4}{\omega^2}} \, \right)\nonumber \\ 
&& \times \delta_{j,j'} \, \delta_{j,N} \,  \delta_{\nu,\nu'}
\end{eqnarray}
on site $N$ of each wire $\nu$. The conductance of the infinite system
Eq.~(\ref{fullH}) can then be calculated considering a finite system
of $MN+1$ lattice sites.  

The flow equations of the
matrix elements with $j,j \pm 1 \in[1,N]$ are
\begin{eqnarray}
\label{flowsigma1}
 \frac{\partial}{\partial\Lambda}  \Sigma^{\nu,\Lambda}_{j,j} &=&
 - \frac{1}{2\pi}  \sum_{\omega = \pm\Lambda} \sum_{r = \pm 1}
U^{\Lambda}_{j,j+r}  \nonumber \\ && \times 
\left<j+r,\nu\right| {\mathcal G}^{\Lambda}(i\omega) \left|
  j+r,\nu\right> \, ,
\\
\label{flowsigma2}
 \frac{\partial}{\partial\Lambda}  \Sigma^{\nu,\Lambda}_{j,j \pm 1} & =&
    \frac{U^{\Lambda}_{j,j\pm 1}}{2\pi} \sum_{\omega = \pm\Lambda}
  \left<j,\nu\right| {\mathcal G}^{\Lambda}(i\omega) \left|
    j\pm1,\nu\right> \, , \;\;\; \;\;\; 
\end{eqnarray}
with the propagator 
\begin{equation}
 {\mathcal G}^{\Lambda}(i\omega) = 
 \left[ {\mathcal G}_0^{-1}(i\omega) - V_{\rm lead}(i\omega) 
-\Sigma^{\Lambda} \right]^{-1} \; ,
\end{equation}
which is a $(MN+1) \times (MN+1)$-matrix, and  
the initial condition 
\begin{eqnarray}
\label{incond}
\Sigma^{\nu,\infty}_{j,j} = 0 = \Sigma^{\nu, \infty}_{j,j\pm 1} \; .
\end{eqnarray}
We introduced the notation
\begin{equation*}
\Sigma^{\nu,\Lambda}_{j,j'} = \left<j,\nu\right| \Sigma^{\Lambda} 
\left| j',\nu\right> \; . 
\end{equation*}  
The matrix elements of the self-energy between the first sites of the wire
and the dot site vanish as there is no interaction across these bonds.

In a numerical solution of Eqs.~(\ref{flowsigma1}) and 
(\ref{flowsigma2}) the flow starts at a 
large finite initial cutoff $\Lambda_0$. One has to take into account 
that, due to the slow decay of the right-hand side (rhs) of the flow equation 
for $\Sigma^{\Lambda}$, the integration from $\Lambda=\infty$ to
$\Lambda=\Lambda_0$ yields a contribution which does not vanish for
$\Lambda_0 \to \infty$, but rather tends to a finite
constant.\cite{Sabine} The resulting initial condition at 
$\Lambda = \Lambda_0 \to \infty$ reads
\begin{eqnarray}
\label{incondfinal}
\Sigma^{\nu,\Lambda_0}_{j,j} & = &  \left( U_{j-1,j} +
  U_{j,j+1} \right)/2  \nonumber \\ 
\Sigma^{\nu,\Lambda_0}_{j,j\pm 1}  & = & 0 \; .
\end{eqnarray}
As we show in the next subsection the inversion of the $(MN+1)
\times (MN+1)$-matrix on the rhs of Eqs.~(\ref{flowsigma1}) 
and (\ref{flowsigma2})
can be reduced to the inversion of $M$ matrices of size $N \times N$. 

At the end of the fRG flow the self-energy $\Sigma^{\Lambda=0}$
presents an approximation for the exact self-energy and
will be denoted by $\Sigma$ in the following. 
In a last step to obtain the Green
function ${\mathcal G}(z)$ which enters Eq.~(\ref{conduct}) we have
to invert the matrix ${\mathcal G}_0^{-1}(z) -V_{\rm lead}(z) - \Sigma$, 
i.e.\ solve the single-particle scattering problem 
with $\Sigma$ and the unrenormalized junction
as potentials. For $U \neq 0$, due to the 
fRG procedure, $\Sigma$ and thus the Green function as well as 
the conductance explicitly depend on the number of sites in  
the interacting part of the wire $N$ and the number of wires $M$. 
These dependences are suppressed in the notation we use.  
Similar to the case of a localized impurity in a single infinite
wire\cite{MatveevGlazman,VM1,Sabine} 
in each wire $\nu$ the real space matrix elements 
$\Sigma^{\nu}_{j,j}$ 
and $\Sigma^{\nu}_{j,j\pm1}$ have an important spatial
dependence. They show a long-range oscillatory behavior around an
average value with an
amplitude which slowly decays with increasing distance from the
junction. The scattering off this potential leads to the power-law
scaling of the conductance.   

Considering different type of geometries of inhomogeneous TLLs 
(single impurity, double barrier, triangular Y-junction with 
a magnetic flux) it
was earlier shown that the above approximation scheme leads to
accurate results for weak to intermediate interactions such that $1/2
\leq K \leq 3/2$.\cite{VM1,VM5,VM4,Sabine,VMresotun,Enss04,Xavier} 
In particular, in
cases where exact expressions for scaling exponents are known from
field theoretical models they are reproduced to leading order in the
interaction.  

Instead of analyzing the scaling of the conductance 
as a function of $\Lambda$ for a given set 
of junction parameters as well as fixed $U$ and $N$, we 
always integrate the flow equations down to $\Lambda=0$ and 
use the energy scale
\begin{eqnarray}
\label{levelspacing}
\delta_N = \frac{\pi v_F}{N} 
\end{eqnarray} 
as our scaling variable. It constitutes an infrared cutoff 
of any power-law scaling with interaction dependent 
exponents.\cite{Enss04} This
procedure has the advantage that each value of the scaling
variable corresponds to a physical system.

\subsection{Scattering theory}
\label{scatterscatter}
      
In this section we use single particle scattering theory to
reach two
goals. (i) We are aiming at an expression for the matrix elements of
the $\Lambda$-dependent Green function entering the rhs of
Eqs.~(\ref{flowsigma1}) and (\ref{flowsigma2}) that 
only requires the inversion of $M$
matrices of size $N
\times N$. This way the numerical effort to integrate the flow
equations can considerably be reduced. (ii) Similarly to the case 
of resonant tunneling in a single infinite wire\cite{Enss04} 
we want to derive equations using  Eq.~(\ref{conduct}) in
which the 
effective transmission is expressed
in terms of the diagonal matrix elements of an auxiliary Green 
function at site $1$ of each wire and the dot site Green function. 
We derive two relations of this type that will be helpful to gain a 
deeper understanding of our results. In the case of a symmetric
junction one of them directly leads to our first result for the 
conductance.
  
The Green function  ${\mathcal G}(z)=\left[{\mathcal G}_0^{-1}(z)
 - V_{\rm lead}(z) - \Sigma\right]^{-1}$ can 
be understood as the resolvent matrix of an effective single-particle 
Hamiltonian $h_{1p}(z)$ with a Hilbert space of size $NM+1$. A 
single-particle basis is given by the states $\{\left| j,\nu \right>,
\left|d\right>\}$, where $\left|d\right>$ denotes the Wannier state
centered on the dot site. The single-particle version of the junction
Hamiltonian Eq.~(\ref{dotH}) is 
\begin{equation}
\label{dotHsingle}
 h_{\rm junc} = -\sum_{\nu=1}^{M} t_{\nu} 
\left( \left| 1,\nu \right> \left< d \right| +  \mbox{H.c.}  \right) + 
V   \left| d \right>  \left< d \right| \, .
\end{equation}
The resolvent can be decomposed as 
\begin{equation}
\label{decom}
 {\mathcal G}(z) = {\mathcal G}_{\rm dc}(z)+{\mathcal G}_{\rm dc}(z)
h_{\rm junc}{\mathcal G}(z)
\end{equation}
with ${\mathcal G}_{\rm dc}(z)$ being the resolvent of the disconnected wires
\begin{equation}
\label{discon}
{\mathcal G}_{\rm dc}(z) = \left[ z - h_{1p}^{0}(z) \right]^{-1} \; .
\end{equation}
The Hamiltonian $h_{1p}^{0}$ follows from $h_{1p}$ after taking
$t_{\nu} =0$ for all $\nu$.
Applying the projector $P= \sum_{\nu=1}^{M}P_{\nu} $ with 
$P_{\nu}=\sum_{j=1}^{N} \pj{j,\nu}$, to the left and right in 
Eq.~(\ref{decom}) one obtains
\begin{eqnarray}
\label{projg}
P {\mathcal G}(z) P & = & \sum_{\nu}{\mathcal G}^\nu_{\rm dc} + 
{\mathcal G}_d(z) \nonumber \\ && 
\times \sum_{\nu,\nu'} \, t_{\nu} t_{\nu'}{\mathcal G}^{\nu}_{\rm dc}(z)
\ket{1,\nu}\bra{1,\nu'}{\mathcal G}^{\nu'}_{\rm dc}(z)
\end{eqnarray} 
with
\begin{eqnarray}
\label{dG}
{\mathcal G}_d(z)=
\left[ z-V-\sum_{\nu}t_{\nu}^2\bra{1,\nu}{\mathcal G}^{\nu}_{\rm dc}(z)
\ket{1,\nu}\right]^{-1} 
\end{eqnarray} 
and 
\begin{eqnarray}
\label{disconproj}
{\mathcal G}^\nu_{\rm dc} = P_{\nu} {\mathcal G}_{\rm dc}(z) P_{\nu} 
= \left[ z P_{\nu} 
-  P_{\nu} h_{1p}^0(z) P_{\nu}\right]^{-1}  .
\end{eqnarray} 
Calculating ${\mathcal G}^\nu_{\rm dc}$ for fixed $\nu$ requires the 
inversion of a $N \times N$ matrix. 
 
The steps leading to Eq.~(\ref{projg}) can also be performed at any
finite cutoff scale $\Lambda$. To determine the matrix elements of 
${\mathcal G}^{\Lambda}$ entering the rhs of the flow equations
for the self-energy (\ref{flowsigma1}) and (\ref{flowsigma2}) 
thus requires the knowledge 
of the tridiagonal parts of the inverse of $M$ tridiagonal $N \times
N$-matrices $z P_{\nu} -  P_{\nu} h_{1p}^0(z) P_{\nu}$ and 
a single column of each of these matrices. Numerically both can be 
computed in ${\mathcal O}(N)$ time.\cite{Sabine} We can therefore
easily treat a fairly large number $M$ of wires each of up 
to $N=10^7$ lattice sites with non-vanishing nearest-neighbor interaction. 

Along the lines of Eqs.~(19) to (24) of Ref.~\onlinecite{Enss04} it is
straightforward to derive the relations
\begin{equation}
\label{taux1}
|t_{\nu,\nu'}(\varepsilon)|^2=4 \Delta_\nu(\varepsilon)\Delta_{\nu'} 
(\varepsilon) |{\mathcal G}_{d}(\varepsilon+i0)|^2.
\end{equation}
and 
\begin{equation}
\label{taux2}
\left|t_{\nu,\nu'}(\varepsilon)\right|^2=
\frac{4 \, \Delta_{\nu}(\varepsilon)\Delta_{\nu'}(\varepsilon)} 
{\left[ \varepsilon - V -\sum_{\nu''}\Omega_{\nu''}(\varepsilon)\right]^2+
\left[\sum_{\nu''}\Delta_{\nu''}(\varepsilon)\right]^2} \, ,
\end{equation} 
with real functions $\Omega_{\nu}(\varepsilon)$ and 
$\Delta_{\nu}(\varepsilon)$ given by
\begin{eqnarray}
\label{aux}
\Omega_{\nu}(\varepsilon)-i\Delta_{\nu}(\varepsilon) =  
t_{\nu}^2\bra{1,\nu}{\mathcal G}^{\nu}_{\rm dc}(\varepsilon+i0)\ket{1,\nu} \, .
\end{eqnarray}
Eqs.~(\ref{taux1}) and (\ref{taux2}) are the expressions for the effective
transmission which can be used instead of Eq.~(\ref{conduct}). 
Later we make extensive use of these relations.

For $V=0$ due to 
particle-hole symmetry $\Omega_\nu (0)=0$ at  
arbitrary $U$. If we furthermore consider a 
symmetric junction  with $t_1=t_2=\ldots =t_M = \tilde t$ 
($Z_M$-symmetric junction) all $\Delta_\nu(0)$ are 
equal and Eq.~(\ref{taux2}) simplifies to  
$\left|t_{\nu,\nu'}(0)\right|^2 = 4/M^2 $
independently of $U$ and $N$. The resulting conductance 
\begin{equation}
\label{maxconduct}
G_{\nu,\nu'} = \frac{e^2}{h} \, \frac{4}{M^2} \; ,
\end{equation} 
with $\nu, \nu' = 1, \ldots, M$ and $\nu \neq \nu'$, 
not only follows in the case of a junction with 
$\tilde t=1$, but for all $\tilde t > 0$. This can be 
explained as a resonance phenomenon. 
Within our approximation scheme 
we thus obtained our first result for
the conductance of an interacting system. 
As $G_{\nu,\nu'}$ is independent of $\delta_N$ we identify 
the above case as a FP. More precisely it corresponds to a
one-parameter line of FPs as the hopping $\tilde t$ can be varied
freely. When considering this case we most of the time leave the
dependence on $\tilde t$ implicit and refer to it as a FP. 
For $M$ non-interacting wires $ \frac{e^2}{h} \, \frac{4}{M^2}$ is the 
conductance maximally allowed by the unitarity requirement of the 
$S$-matrix. We thus denote this FP as the ``perfect junction'' FP.
As our approximation is correct to order $U$ we can
conclude that if there is any interaction dependent correction to 
Eq.~(\ref{maxconduct}) it must be at least of order $U^2$.

To gain additional analytical insight we next study the one-particle 
spectral function for a symmetric junction of $M$ wires. From the
spectral function the $\delta_N$-dependence of the conductance from one of
the equivalent wires to an additional wire that is only weakly
coupled to the junction can be deduced. 

\section{The Hartree-Fock approximation for symmetric junctions with
  $\bm{V=0}$}
\label{HF}

The one-particle spectral function $\rho_{\rm obc}$ of a TLL with 
an open boundary, taken at the chemical potential and close to the 
boundary, shows power-law scaling as a function of $\delta_N$ with 
the exponent $\alpha_B = 1/K-1$.\cite{KaneFisher,FG,VM1,Sabine} 
Remarkably, for $U > 0$ this behavior of the spectral function 
can already be obtained from the non-self-consistent HFA, with a scaling 
exponent\cite{OBCVM} 
\begin{eqnarray}
\label{HFboundaryexp}
\alpha_B^{\rm HF} =  \frac{U/\pi}{1+U/\pi} \; .
\end{eqnarray}
Using Eq.~(\ref{Kpert}) it is straightforward to show that 
$\alpha_B^{\rm HF}$ and $\alpha_B$ agree to leading order in the
interaction $U$. The appearance of power-law scaling within the 
HFA can be traced back to the spatial dependence of the self-energy.
Similar to the fRG approximation of the 
self-energy\cite{VM1,Sabine} the HFA self-energy
shows a long-range oscillatory dependence on $j$ that implies
power-law scaling of the spectral weight (see below). In contrast to
the boundary problem in the single impurity problem the HFA
does not reproduce the correct exponent even to leading order in $U$
as the essential RG flow of the impurity is not included.\cite{VM1} 
One can expect that the HFA leads to meaningful results in all 
models of inhomogeneous TLLs with repulsive interaction in which such 
a flow is unimportant. 
For attractive interactions the  HFA can not be used even for 
the boundary problem as $\alpha_B^{\rm HF} \propto |U|>0$, 
while the exact exponent $\alpha_B$ is negative. We note in passing 
that for inhomogeneous TLL the application of 
the self-consistent HFA leads to unphysical results. As an artifact, 
the iteration of the Hartree-Fock equations generates a groundstate 
with charge-density wave order.  

The above insights motivate us to study the spectral function 
$\rho_1$ on lattice site 1 (of one of the wires) for a symmetric 
junction of $M$ wires, with $t_{\nu}=\tilde t$ for $\nu = 1,\ldots, M$ 
and $V=0$ using the HFA. We show analytically that at the chemical
potential 
\begin{eqnarray}
\label{rho1scal}
\rho_1 \propto \delta_N^{-\alpha^{\rm HF}_M} \; ,
\end{eqnarray}
with a scaling exponent $\alpha^{\rm HF}_M$ that depends on $M$ and 
$U$. In the following we refer to the spectral function evaluated 
at the chemical potential $\mu=0$ as ``the spectral weight''.

In the HFA and for $t_{\nu}=\tilde t$ as well as $V=0$ 
the Hamiltonian (\ref{fullH}) reads
\begin{align}
H_{\rm HF}=&-\sum_{\nu=1}^{M}\sum_{j=1}^{\infty} t(j)\big( \,
c^{\dag}_{j+1,\nu} c_{j,\nu}^{\phantom\dag} + c^{\dag}_{j,\nu} 
\, c_{j+1,\nu}^{\phantom\dag}
  \, \big) \nonumber \\
  & -\tilde{t}\sum_{\nu=1}^{M}\big( \,c^{\dag}_{1,\nu}
  d^{\phantom\dag} + d^{\dag}\, c_{1,\nu}^{\phantom\dag} \, \big) ,
\end{align}
where 
\begin{eqnarray}
\label{tjdef}
t(j)=  \left\{  \begin{array}{ll}  
1+U_{j,j+1} \langle c^{\dag}_{j+1,\nu}c^{\phantom\dag}_{j,\nu} 
\rangle_0 & \; \mbox{for} \,  1\leq j < N \\
1 & \;  \mbox{otherwise} \; ,
  \end{array} \right. 
\end{eqnarray}
with the HFA self-energy $\Sigma^{\rm HF}_{j,j+1} =  U_{j,j+1} 
\langle c^{\dag}_{j+1,\nu}c^{\phantom\dag}_{j,\nu} \rangle_0$. 
The expectation value $\langle \ldots \rangle_0$ is taken with the 
many-body groundstate of $H_{\rm kin}+H_{\rm junc}(V=0)=H_{\rm
  HF}(U=0)$. As we
shifted the density by its average value the Hartree term 
vanishes. 

The normalized single-particle eigenstates 
$ \ket{\Psi_{\varepsilon},l}$ 
of the one-particle version $h_{\rm HF}$ 
of the Hartree-Fock Hamiltonian 
can be classified according to their behavior
in $2\pi/M$-rotations ($Z_M$-symmetric case) which leads to the 
expansion
\begin{equation}
  \label{Psi}
  \ket{\Psi_{\varepsilon},l}=\delta_{l,0}a_0(\varepsilon)\ket{d}+
  \sum_{\nu=1}^{M} \sum_{j=1}^{\infty} e^{i\frac{2\pi l}{M} \nu}
  a_{j}^{(l)}(\varepsilon)  \ket{j,\nu}\; ,
\end{equation}
with coefficients $a_{j}^{(l)}(\varepsilon)$ and $a_0(\varepsilon)$.
For odd values of $M$ one has $l=0,\pm1, \ldots ,\pm (M-1)/2$, while for
even values $l=0,\pm1, \ldots ,\pm M/2-1,M/2$. 

\subsection{The eigenstates for $\bm{U =0}$}

In a first step we determine the eigenstates of the non-interacting system.  
The energies are $\varepsilon= \varepsilon(k)=-2\cos(k)$. For fixed
energy the wave number is thus given by 
$k(\varepsilon) = \arccos(-\varepsilon/2)$.   
Using Eq.~(\ref{Psi}) 
\begin{eqnarray*}
h_{\rm HF}(U=0) \ket{\Psi_{\ep},l}_0 =\ep\ket{\Psi_{\ep},l}_0
\end{eqnarray*}
yields 
\begin{eqnarray}
  \label{U01}
  \varepsilon a_{j}^{(l)}(\ep) &=&-a_{j+1}^{(l)}(\ep)
-a_{j-1}^{(l)}(\ep)(1-\delta_{j,1}) \nonumber \\  
&& -\tilde{t}\delta_{l,0}\delta_{j,1}   a_0(\ep) 
\end{eqnarray}
for the coefficients of $ \ket{\Psi_{\ep},l}_0$. 
For $l=0$ there is an additional equation
\begin{equation}
  \label{U02}
  \varepsilon a_0(\ep)=-M\tilde{t}a_1^{(0)}(\ep) \, .
\end{equation}
The solution  of Eqs.~(\ref{U01}) and (\ref{U02}) is  
\begin{equation}
\label{form1}
  a_{j}^{(l)}(\varepsilon)=A(M,\ep)\sin \left\{ j\,k(\varepsilon)+
  \delta_l\left[k(\varepsilon)\right] \right\} 
\end{equation}
and
\begin{equation}
\label{c_0eq}
a_0(\ep)=A(M,\ep)\sin \left\{ \delta_0\left[k(\varepsilon)\right] 
\right\}/\tilde{t} \; ,
\end{equation}
with the phase shifts 
\begin{eqnarray}
\label{phaseshifts}
  \delta_l(k) = 
 \left\{  \begin{array}{ll}  
 \arctan \left\{ \frac{M \tilde{t}^2}
  {2- M \tilde{t}^2}\tan (k) \right\}     
  \; & \mbox{for} \; l=0 \\
    0
    \; & \mbox{otherwise} \; , 
  \end{array} \right.
\end{eqnarray}
the normalization factor
\begin{equation}
\label{normconstant}
A(M,\ep)=\sqrt{\frac{2}{\pi\,M\,v(\varepsilon)}} \; ,
\end{equation}
and the velocity $v(\varepsilon)= 2
\left|\sin\left[k\left(\varepsilon\right)\right]\right|$.

The resulting single-particle states $\ket{\Psi_{\ep},l}_0$ can then
be used to calculate the groundstate expectation value that enters the HFA
Hamiltonian
\begin{eqnarray}
\langle c^{\dag}_{j+1,\nu}c^{\phantom\dag}_{j,\nu} 
\rangle_0  & = & \sum_l \int_{-B/2}^{\mu} d \, \ep \, a_{j+1}^{(l)}(\ep) 
\, a_{j}^{(l)}(\ep)  \nonumber \\
& = & \frac{1}{\pi M} \sum_l \int_0^{k_F}  d\,k  \left\{ \cos[k]
 \right. \nonumber \\  &&  \left. - \cos
  \left[k (2j+1) + 2 \delta_l(k) \right] \right\} \; ,
\label{expectval1}
\end{eqnarray}  
with the Fermi wave number $k_F=\pi/2$ and band width $B=4$. 
As described below to determine the scaling of the spectral weight of 
the interacting system  for $\delta_N \to 0$ we only need to know 
the behavior of this expectation value for $1 \ll j < N$
\begin{eqnarray*}
\label{expectval2}
\langle c^{\dag}_{j+1,\nu}c^{\phantom\dag}_{j,\nu} 
\rangle_0  = \frac{1}{\pi} \left[ 1 - 
\frac{(-1)^j}{2j+1} \left(1-\frac{2}{M}\right)  +
{\mathcal O}\left(j^{-2}\right)\right].
\end{eqnarray*} 
This asymptotic behavior can be obtained from 
Eq.~(\ref{expectval1}) using integration by 
parts and $\delta_0(k_F) = \pm \pi/2$. For $1 \leq j \leq N-j_s$
the interaction $U_{j,j+1}$ Eq.~(\ref{shape}) takes its bulk value 
$U$ and it follows that  
\begin{eqnarray}
\label{efft}
t(j) & = & 1+\frac{U}{\pi} - \frac{U}{\pi} \frac{(-1)^j}{2j+1}
\left(1-\frac{2}{M}\right) \nonumber \\
& = & \left( 1+\frac{U}{\pi} \right) \left[ 1 - \alpha_B^{\rm HF} 
\frac{(-1)^j}{2j+1} \left(1-\frac{2}{M}\right)  \right] \; ,
\end{eqnarray} 
for $1 \ll j \leq N-j_s$.

\subsection{Spectral weight for $\bm{U \neq 0}$}

Within the HFA the spectral weight on the first site of one of the 
equivalent legs is determined by the amplitudes of the $\varepsilon=0$ 
eigenstates $\ket{\Psi_{\ep=0},l}$ of $h_{\rm HF}$
\begin{eqnarray}
\label{rhowf}
\rho_1  \propto \sum_l \left| \left< 1, \nu
  \right. \ket{\Psi_{\ep=0},l} \right|^2  =  \sum_l \left|
  a_1^{(l)}(0) \right|^2 \; .
\end{eqnarray}
To avoid proliferation of symbols or indices the expansion
coefficients of the interacting HFA eigenstates are denoted by the
same symbols $a_j^{(l)}$ and $a_0$ as the coefficients of the 
non-interacting eigenstates in the last subsection.
With the expansion Eq.~(\ref{Psi}) the stationary Schr\"odinger equation 
\begin{eqnarray*}
h_{\rm HF} \ket{\Psi_{\ep=0},l} = 0 
\end{eqnarray*} 
leads to coupled equations for the coefficients $a_j^{(l)}(0)$. The $l=0$
equation (\ref{U02}) also holds for $U \neq 0$ which for $\varepsilon
=0$ leads to $a_1^{(0)}(0) = 0$. Only the $l=0$ eigenstate
$\ket{\Psi_{\ep=0},0}$ has a non-vanishing amplitude on the dot site
which implies that the $a_1^{(l)}(0)$ for $l\neq 0$ can be calculated as
for a semi-infinite chain. For 
$j\geq 2$ and $\varepsilon=0$ the Schr\"odinger equation gives
\begin{eqnarray*}
\label{iteq}
\bra{j,\nu} h_{\rm HF} \ket{\Psi_{\ep=0},l} = 0 = -t(j) a_{j+1}^{(l)}(0) -
t(j-1) a_{j-1}^{(l)}(0) \; . 
\end{eqnarray*} 
This relation can be solved iteratively leading to 
\begin{eqnarray}
\label{soliteq}
a_{2j+1}^{(l)} = (-1)^j \, a_1^{(l)}(0) \, \prod_{i=1}^{j}
\frac{t(2i-1)}{t(2i)} \; .
\end{eqnarray} 
Because of  
\begin{eqnarray*}
\bra{1,\nu} h_{\rm HF} \ket{\Psi_{\ep=0},l} = 0 = -t(1) a_2^{(l)}(0) \; ,
\end{eqnarray*}
$a_j^{(l)}(0)=0$ for all even $j$. 

Without loss of generality
we now consider the case of even $N$.
As the interaction $U_{j,j+1}$ vanishes for $j \geq N$, $t(2i-1) =
t(2i)$ for $i \geq N/2+1$ and  $\left| a_{2j+1}^{(l)} \right|$ 
is independent of $j$ for $j \geq N/2+1$. Together with the 
asymptotic scattering state form of the
$a_{j}^{(l)}(0)$ Eq.~(\ref{form1}) this implies that with respect to the
$\delta_N$-dependence we find 
\begin{eqnarray}
\label{a1form1}
\rho_1 \propto  
\sum_l \left| a_{1}^{(l)} \right|^2 \propto \left| \prod_{i=1}^{N/2}
\frac{t(2i)}{t(2i-1)} \right|^2 \; .
\end{eqnarray}
We next evaluate the rhs of this expression for large $N$. We
do this separately for the numerator and denominator.  
Using Eq.~(\ref{efft}) one obtains
\begin{eqnarray*}
\label{num}
&& \ln \left[\prod_{i=1}^{N/2} t(2i) \right]  =  \sum_{i=1}^{N/2} \ln
\left[ t(2i) \right] \nonumber \\ & = & \frac{N}{2} 
\ln\left(1+\frac{U}{\pi}\right) -\frac{1}{4} \,  
\alpha_B^{\rm HF} \left(1-\frac{2}{M} \right) \, \ln N + {\mathcal
  O}(N^0) \; .
\end{eqnarray*}
The first two terms follow from the factors of the product in which 
$t(2i)$ can be replaced by Eq.~(\ref{efft}). The remaining factors lead
to the last summand of order $N^0$.   
Similarly one gets 
\begin{eqnarray*}
\label{denom}
&& \ln \left[\prod_{i=1}^{N/2} t(2i-1) \right] = \sum_{i=1}^{N/2} \ln
\left[ t(2i-1) \right] \nonumber \\ 
& = & \frac{N}{2}  \ln\left(1+\frac{U}{\pi}\right) +\frac{1}{4} \,  
\alpha_B^{\rm HF} \left(1-\frac{2}{M} \right) \, \ln N + {\mathcal
  O}(N^0) \; .
\end{eqnarray*}
Combining these two asymptotic expressions leads to the power-law
scaling of the spectral weight on the first lattice site of each wire 
\begin{eqnarray}
\label{specasymp}
\rho_1 \propto  N^{\alpha^{\rm HF}_M} \propto
\delta_N^{-\alpha^{\rm HF}_M} \; , 
\end{eqnarray}
with 
\begin{eqnarray}
\label{expdef} 
\alpha^{\rm HF}_M = \alpha_B^{\rm HF} \left( \frac{2}{M}-1 
\right) = \frac{U/\pi}{1+U/\pi} \left( \frac{2}{M}-1 
\right) \; . 
\end{eqnarray}
This derivation shows that the power-law scaling of the
spectral weight directly follows from the long-range spatial
oscillations of the self-energy.

Using Eqs.~(\ref{projg}) and (\ref{dG}) it is straightforward to show 
that  for a symmetric junction  
the spectral weight on the first lattice site  $\rho_1$ and the 
spectral weight on the dot site $\rho_d$ are inversely proportional to 
each other. It thus follows that 
\begin{eqnarray}
\label{rhodotscal}
\rho_d \propto \delta_N^{\alpha^{\rm HF}_M} \; .
\end{eqnarray}
For $M=1$ the dot site is the last
site of a semi-infinite chain with open boundary conditions and 
$\alpha^{\rm HF}_1$ agrees with $\alpha_B^{\rm HF}$. For $M=2$ 
the dot site corresponds to a bulk lattice site of a
homogeneous TLL. As long as $V=0$ this holds even for $\tilde t \neq
1$ due to a resonance. In a homogeneous TLL the spectral weight scales as 
$\delta_N^{\alpha}$, with $\alpha=(K^{-1}+K-2)/2$.\cite{KS} Replacing 
$K$ in this expression by the leading order term Eq.~(\ref{Kpert}) 
it follows that $\alpha = \mathcal{O}(U^2)$. The HFA exponent can 
only expected to be  correct to leading order in $U$. Consistently we 
find $\alpha^{\rm HF}_2=0$. At $M=2$ and $U>0$ the sign of 
$\alpha^{\rm HF}_M$ changes from $\alpha^{\rm HF}_M >0$ for $M=1$ to 
$\alpha^{\rm HF}_M <0$ for $M \geq 3$. 

\subsection{The conductance across a weak link}

We next consider a junction of $M$ wires, $M-1$ of them with hopping
$\tilde t$ from the first lattice sites of the wires to the dot 
site, while the leg with $\nu=M$ is coupled by the hopping 
amplitude $t_{M} \ll \tilde t$.
The HFA result for the spectral weight can be used to determine the 
scaling exponent of the conductance from one of the $M-1$ equivalent wires 
to the wire $M$. Applying Fermi's golden rule one can argue that 
in this tunneling limit the scaling of the conductance across the weak 
link is determined by the product of $\rho_{\rm obc}$ and $\rho_d$
leading to 
\begin{eqnarray}
\label{scalsymmweak}
G_{\nu,M} \propto \delta_N^{\gamma_1^{\rm HF}(M)}
\end{eqnarray}
for $\nu=1,\ldots,M-1$, with 
\begin{eqnarray}
\label{gamma1HF}
\gamma_1^{\rm HF}(M) = 2 \alpha_B^{\rm HF}/(M-1) \; .
\end{eqnarray}
This constitutes our second result for 
the transport through a dot junction.  
For $M=2$ the junction problem is equivalent to the single impurity 
problem in the limit of a weak link (strong impurity) which is
characterized by the exponent $2
\alpha_B$.\cite{KaneFisher} We reproduce this result to leading order
in $U$.
Anticipating  the RG language Eq.~(\ref{scalsymmweak}) indicates 
that for $U>0$ a 
weak link of a symmetric $(M-1)$-leg junction to an additional wire is 
an irrelevant perturbation. The exponent $ \gamma_1^{\rm HF}(M) > 0$, 
i.e. $G_{\nu,M} \to 0$ and with decreasing $\delta_N$ the system 
``flows back'' to the ``perfect junction'' FP of 
the $(M-1)$-wire system with the conductance Eq.~(\ref{maxconduct}). 

Without using the golden rule like argument and applying the fRG we 
next numerically  confirm the power-law scaling of $G_{\nu,M}$
as well as the RG interpretation of the HFA results. Going beyond the
symmetric junction with $V=0$ and an additional weak link, we 
study the conductance for general junction parameters. 
The fRG can also be used for $U<0$.

\section{fRG Results}
\label{fRGresults}

In this section we present the results for the FPs and the scaling of the
conductance obtained from numerically solving the fRG
flow equations of Sect.~\ref{method} for $U \neq 0$. In subsections 
\ref{sub1} and \ref{sub2} we investigate 
two specific classes of junction parameters. The results from these
cases can be combined and lead to the comprehensive picture for 
arbitrary junction parameters presented in subsection \ref{sub3}.

\subsection{A symmetric junction with one modified link} 
\label{sub1}

We here consider a junction of $M \geq 2$ wires, $M-1$ with hopping $\tilde
t$ (not necessarily $\tilde t=1$) between the first site of the wire 
and the dot, while the 
additional one has hopping $t_{M}\neq \tilde{t}$. The 
dot site energy $V$ we set to zero and the real part $\Omega_{\nu}(0)$
of the auxiliary Green function Eq.~(\ref{aux}) vanishes. In addition 
$\Delta_\nu(0) = \Delta_{\nu'}(0) = \Delta$ for all $\nu, \nu' \leq M-1$.    
Two cases  can be distinguished.

\subsubsection{A weak link}
\label{subsub1}

\begin{figure}[bth]
\begin{center}
\includegraphics[width=0.4\textwidth,clip]{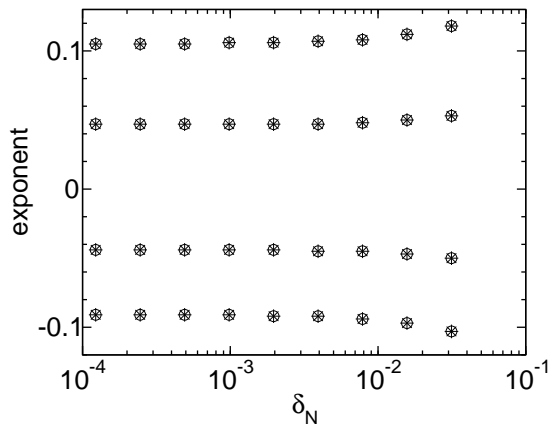}
\end{center}
\caption{Effective exponents of 
$\frac{4}{(M-1)^2}-|t_{\nu,\nu'}|^2$ (circles) and $|t_{\nu,M}|^2$ (stars) 
as a function of $\delta_N$ for $M=5$, $\tilde t=1$, $t_{M}=10^{-3}$,
and different values of $U=-1,-0.5,0.5,1$ from top to bottom. 
On the scale of the plot the results obtained from
$\frac{4}{(M-1)^2}-|t_{\nu,\nu'}|^2$ and $|t_{\nu,M}|^2$
are indistinguishable.
The scaling exponent is read-off when a plateau is reached for 
$\delta_N < 5 \cdot 10^{-4}$, which roughly corresponds to $N>10^4$.}
\label{f:scaling}
\end{figure}

The first one is the weak link situation already treated by the HFA
with $\tau = t_{M}/\tilde{t} \ll 1$. We thus slightly perturb the
``perfect junction'' FP of a $(M-1)$-wire system in a specific way. 
As expected we numerically find that the effective transmission between 
the $M-1$ legs with $\tilde t$ is close to the perfect transmission
$4/(M-1)^2$. For $U>0$ and decreasing $\delta_N$ the fRG data for 
$|t_{\nu,\nu'}|^2$ approach this value, while they leave it for
$U<0$. At $U=0$, $|t_{\nu,\nu'}|^2$ is independent of $\delta_N$. 
Similarly $|t_{\nu,M}|^2 \to 0$ for $U>0$, while it 
takes a small increasing value for $U<0$. We thus analyze
the power-law scaling of $|t_{\nu,\nu'}|^2$ and $|t_{\nu,M}|^2$ as a
function of $\delta_N$ with respect to $4/(M-1)^2$  and $0$,
respectively. The effective exponents as a function of $\delta_N$ 
obtained by taking the log-derivative of the fRG data calculated for 
$M=5$, $\tilde t=1$, $t_{M}=10^{-3}$, and different $U$ is shown in 
Fig.~\ref{f:scaling} on a log-linear scale. For small $\delta_N$ and
all $U$ both exponents approach the same $U$-dependent plateau value, 
which is our fRG approximation for the scaling 
exponent. Even for larger arguments the two $\delta_N$-dependent
effective exponents are indistinguishable on the
scale of the plot.

\begin{figure}[hbt]
\begin{center}
\includegraphics[width=0.4\textwidth,clip]{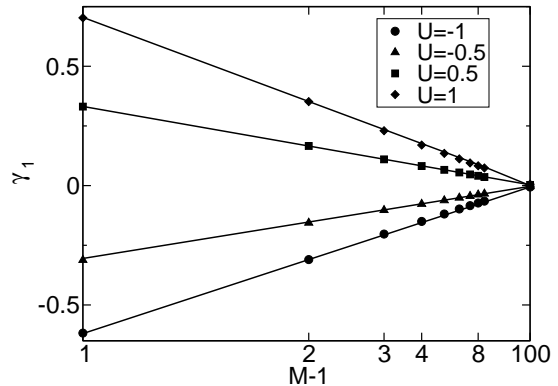}
\end{center}
\caption{Scaling exponent of the conductance for a weak link as a
  function of $M-1$ (symbols) for different $U$. Note the reciprocal scale of 
  the $x$-axis. The lines show $\gamma_1(M)=\beta_s/(M-1)$.}
\label{f:low_link}
\end{figure}

The $M$-dependence of the scaling exponent for different $U$ is shown in
Fig.~\ref{f:low_link} on a reciprocal-linear scale (symbols). The data can
be fitted by (lines)
\begin{eqnarray}
\label{gamma1}
\gamma_1(M)=\beta_s /(M-1) \, ,
\end{eqnarray}
with $\beta_s$ being the $U$-dependent fRG approximation for $2
\alpha_B=2(1/K-1)$ obtained in Ref.~\onlinecite{Enss04}
for a weak link (strong impurity) in an infinite wire, i.e.~the $M=2$
case. The exponent $\beta_s$ agrees with 
$2\alpha_B$ to leading order in $U$. It has higher 
order corrections which  
bring it close to $2\alpha_B$ even for intermediate $U$. 
For the four interactions $U=-1,-0.5,0.5,1$ 
the value for $\beta_s$ can be read off from Fig.~\ref{f:low_link}. 
At $U=1$ the relative difference between the exact exponent obtained
from Eq.~(\ref{Kpert}) and our
approximation is roughly 5\%. 
A detailed comparison of $\beta_s$ and $2 \alpha_B$ is given 
in Fig.~5 of Ref.~\onlinecite{Enss04}. 
To leading order in $U$ Eq.~(\ref{gamma1}) confirms the result 
deduced from the combined use of the HFA and 
Fermi's golden rule arguments Eq.~(\ref{gamma1HF}). 
For $M=2$, the fRG approximation $\beta_s$ is generically closer 
to the exact exponent $2 \alpha_B$ than $2 \alpha_B^{\rm HF}$.

\begin{figure}[bt]
\begin{center}
\includegraphics[width=0.4\textwidth,clip]{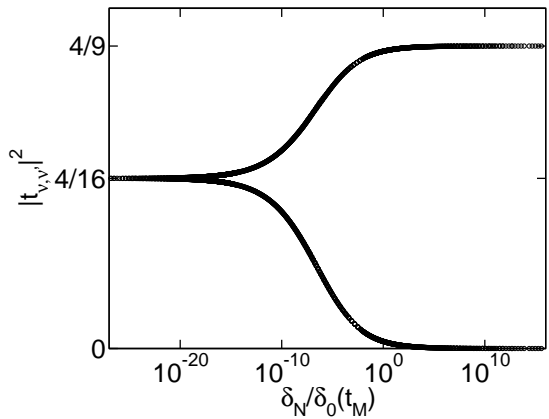}
\end{center}
\caption{One-parameter scaling plot of the effective 
transmissions $|t_{\nu,M}|^2$ (lower curve) and $|t_{\nu,\nu'}|^2$ 
with $\nu,\nu' \leq M-1$ (upper curve) 
for $M=4$, $U=-1$, and $\tilde
t=1$.  The variable is $\delta_N/\delta_0(t_M)$, with a non-universal 
scale $\delta_0(t_M)$. The asymptotic values of the transmission
$4/(M-1)^2$ and $0$ reached for $\delta_N/\delta_0(t_M) 
\to \infty$ as well as  $4/M^2$ reached for $\delta_N/\delta_0(t_M) 
\to 0$ are indicated on the
$y$-axis.}
\label{scalfig}
\end{figure}
 
For $U>0$, $\gamma_1 >0$ and the ``perfect junction'' FP of the 
$(M-1)$-wire system
is stable towards weakly coupling an additional wire, while 
for $U<0$, $\gamma_1 <0$ and the FP is unstable. In the latter case
the system effectively incorporates the weakly coupled leg and flows to 
the ``perfect junction'' FP for $M$ wires. For small to intermediate
$|U|$ and starting close to the ``perfect junction'' FP of the 
$(M-1)$-wire system, i.e.~for $t_M \ll \tilde t$, exponentially small
$\delta_N$ are required to reach the new FP. 
Even though we can treat very large $N$ such small scales are beyond 
the possibilities of our method. Similar to the single impurity
problem\cite{KaneFisher,Fendley,VM4,Enss04} the flow from one   
to the other FP can be shown considering data sets 
$(\delta_N,|t_{\nu,\nu'}|^2)$ obtained for different $t_{M} \in ]0,1[$ 
using a one-parameter scaling ansatz. For fixed $U$ the conductance 
is only a function of $\delta_N/\delta_0(t_M)$, with an appropriately
chosen non-universal energy scale $\delta_0(t_M)$ and the different
data sets can be collapsed on a single curve. This is shown in
Fig.~\ref{scalfig} for the two different
effective transmissions $|t_{\nu,M}|^2$ (lower curve) and
$|t_{\nu,\nu'}|^2$ with $\nu,\nu' \leq M-1$ (upper curve) 
and $M=4$, $U=-1$, $\tilde t=1$ on a log-linear scale. 
It is important to note that the 
effective transmission $4/M^2$ between all legs is not 
achieved by an increase of $t_{M}$. As there is no interaction 
across the links of the dot to the first sites of the wires, 
the hopping across these links is independent  of the RG cutoff 
(here $\delta_N$). The transmission $4/M^2$ follows from the 
build up of a self-energy during the RG flow, that, 
interpreted as an effective 
scattering potential, leads to a resonance at the chemical potential. 
For $\gamma_1 <0$, that is $U<0$, the plateau reached in
Fig.~\ref{f:scaling} does not present the asymptotic behavior for
$\delta_N \to 0$ and considering even smaller $\delta_N$ a deviation
from the plateau value $\gamma_1$ can be observed.

We next show that the scaling behavior of the conductance 
can be understood from the
scaling of the spectral weights on the first site of one of the
equivalent legs and the first site of the additional wire. 
For $\ep=0$ we get from Eq.~(\ref{taux2}) 
\begin{equation}
  \label{e:low_link1}
  |t_{\nu,\nu'}|^2 = \frac{4}{(M-1)^2}
  -\frac{8}{(M-1)^3} D  + {\mathcal O}(\tau^4) \; ,
\end{equation}
for $\nu \neq  \nu'$  and $\nu,\nu' \leq M-1$, as well as 
\begin{equation}
  \label{e:low_link2}
  |t_{\nu,M}|^2 = \frac{4}{(M-1)^2} D
  -\frac{8}{(M-1)^3} D^2 + {\mathcal O}(\tau^6) \, ,
\end{equation}
for $\nu \leq M-1$, with 
\begin{equation}
\label{ratio}
D= \Delta_{M}(0)/\Delta \propto \tau^2 \; .
\end{equation}
For the reflection back into wire $\nu \leq M-1$ it follows
\begin{align}
\label{reflec}
  R_{\nu}&=1-\sum_{\nu'\neq\nu}|t_{\nu,\nu'}|^2\nn \\
  & = \frac{(M-3)^2}{(M-1)^2}+ \frac{4}{(M-1)^2} \left[
    1-\frac{2}{(M-1)}
  \right] D \nonumber \\ &
+ \frac{8}{(M-1)^3} D^2 + {\mathcal O}(\tau^6) \; . 
\end{align}
As discussed in Sect.~\ref{method} the auxiliary Green function is
calculated with a self-energy which has been determined in the
presence of the weak link. Thus $\Delta/\tilde t^2$ and the spectral 
weight $\rho_1$ of a perfect $(M-1)$-leg junction differ 
to order $\tau^2$. If corrections of order $\tau^2$ are neglected 
in $\Delta$ and $\Delta_{M}(0)$, $D$ can be replaced by 
$\rho_{\rm obc}/\rho_1$. We numerically find that for
$\tilde M=1,2,\ldots$ the spectral weight on the first site 
of a perfect $\tilde M$-leg junction scales as 
\begin{eqnarray}
\label{specasympfRG}
\rho_1 \propto \delta_N^{-\alpha^{\rm fRG}_{\tilde M}} \; , 
\end{eqnarray}
with 
\begin{equation}
\label{alphaMfRG}
  \alpha_{\tilde M}^{\rm fRG}=\frac{\beta_s}{2} 
\left( \frac{2}{\tilde M}-1\right) \; ,
\end{equation}
consistent with the derivation using the HFA in Eq.~(\ref{expdef}).
The above replacement thus leads to $D \propto
\delta_N^{\beta_s/(M-1)}$, where we used 
$\rho_{\rm obc} \propto \delta_N^{\beta_s/2}$. 
Inserting this expression in
Eqs.~(\ref{e:low_link1}) and (\ref{e:low_link2}) we reproduce the
results directly obtained from calculating the effective transmission.  
Note that for $M=3$ the second term in Eq.~(\ref{reflec}) cancels and
the reflection scales with an exponent that is twice as large as the
one of the transmission. 
Further down we encounter another case in which the prefactor of the
leading order term vanishes for specific parameters, which leads to a
doubling of the scaling exponent.

In the limit $M \to \infty$ the dot site is coupled to so many 
legs that it creates an infinite barrier. Consistently 
Eq.~(\ref{alphaMfRG}) gives 
$\lim_{M \to \infty} \alpha_M^{\rm fRG} = - \frac{\beta_s}{2}$ and 
the spectral weight on the first sites scales as the weight next to an
open boundary.

\subsubsection{A slightly modified link}

As the second example with a single modified link we consider 
$|t_{M}-\tilde{t}|/\tilde{t}=\tau\ll 1$. We now analyze the scaling
of the effective transmission within the subsystem of the $M-1$
equivalent wires and into the leg with the modified hopping with
respect to the transmission $4/M^2$ of the perfect case. Similarly
to Fig.~\ref{f:scaling} exponents can reliably be extracted for
sufficiently large $N$. 

\begin{figure}[bt]
\begin{center}
\includegraphics[width=0.4\textwidth,clip]{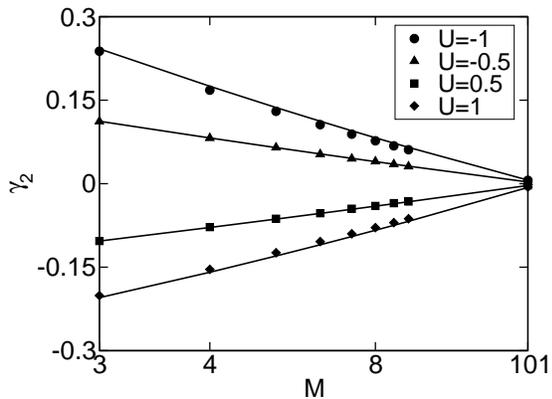}
\end{center}
\caption{Scaling exponent of the conductance for a slightly modified 
  link as a function of $M$ for different $U$. Note the reciprocal scale of 
  the $x$-axis. The lines show the fit
  Eq.~(\ref{gamma2}).}
\label{f:nearly_link}
\end{figure}

For $M \geq 3$ the exponents of
$|t_{\nu,\nu'}|^2$ and $|t_{\nu,M}|^2$, with $\nu,\nu' \in [1,M-1]$
turn out to be equal. For several $U$ the $M$-dependence of the 
scaling exponent is  shown in 
Fig.~\ref{f:nearly_link} on a reciprocal-linear
scale. To first order in $U$ it is given by
$-\beta_s/M$. More accurately the data can be fitted by
\begin{eqnarray}
\label{gamma2}
\gamma_2(M)=-\frac{\beta_s}{M} + 2 \frac{\beta_w+\beta_s}{M^2} \; ,
\end{eqnarray}
with the exponent $\beta_w$ found in Ref.~\onlinecite{Enss04} 
for the scaling of the transmission through a local weak
impurity. To leading order in $U$, 
$\beta_w$ agrees with the exact weak single impurity exponent $2(K-1)$. 
A comparison of $2(K-1)$ and our fRG approximation  
is given in Fig.~7 of Ref.~\onlinecite{Enss04}.

For $U>0$, $\gamma_2<0$ and the ``perfect junction'' FP of the
$M$-wire system is unstable towards changing one of the hopping
amplitudes. For $t_{M} < \tilde t$ the system flows to the 
``perfect junction'' FP  of $M-1$ wires. This follows from the build 
up of a self-energy during the RG flow, that leads to a vanishing 
transmission from wire $\nu \leq M-1$ to wire $M$, but to 
a resonance with perfect transmission between the $M-1$ equivalent legs.  
For $t_{M} > \tilde t$ the flow is to a FP with vanishing
conductance between all wires -- the ``decoupled chain'' FP. 
The vanishing of the conductance follows from the long-range
oscillatory behavior of the self-energy generated in the RG flow. 
In both cases the flow from one to the other FP can again be shown using
a one-parameter scaling ansatz.
For $U<0$, $\gamma_2 > 0$ and the ``perfect junction'' FP of the
$M$-wire system is stable, regardless of the sign of $\tilde t -
t_{M}$.  

\begin{figure*}[tb]
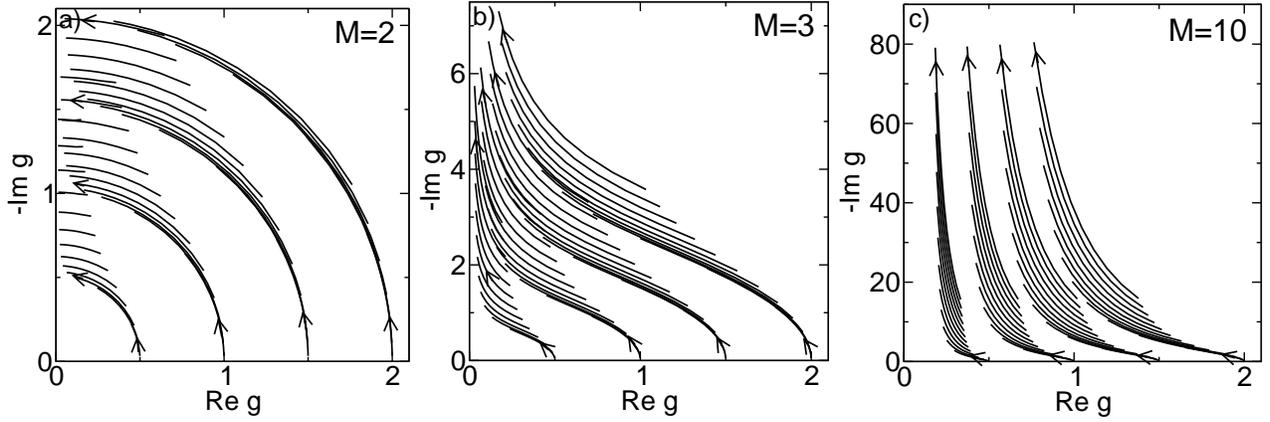

\begin{center}
\includegraphics[width=0.3\textwidth,clip]{w2.eps}
\includegraphics[width=0.305\textwidth,clip]{w3.eps}
\includegraphics[width=0.315\textwidth,clip]{w10.eps}
\end{center}
\caption[]{Flow of $g$ Eq.~(\ref{smallgdef}) for the case of a
  symmetric junction with on-site energy $V=0.5,1,1.5,2$, 
  different $\tilde t \in
  [0.01,0.7]$, $M=2,3,10$ legs, and $U=-1$. The arrows indicate the 
  direction of the flow for $U<0$. For $U>0$ it is reversed. 
\label{planeflow}}
\end{figure*}

For $M=2$ there is no $|t_{\nu,\nu'}|^2$ and $1-|t_{\nu,M}|^2$
scales with $2 \gamma_2(2)$. The appearance of the factor $2$ can be
explained by considering an expansion of $|t_{\nu,M}|^2$  similar to
Eqs.~(\ref{e:low_link1}) and 
(\ref{e:low_link2}) (which is valid for all $M \geq 2$) 
\begin{eqnarray}
  \label{e:nearly_link2}
  |t_{\nu,M}|^2&= &\frac{4}{M^2}+
  \frac{4(D-1)}{M^2}
  \left(
    1-\frac{2}{M}
  \right)  \nonumber \\
  && -\frac{4 (D-1)^2}{M^3} \left(2 - \frac{3}{M} \right) \; ,
\end{eqnarray}
with $D$ as defined in Eq.~(\ref{ratio}). The difference $D-1$ shows
power-law scaling with exponent $\gamma_2(M)$. For $M=2$ 
the prefactor of the leading order term linear in $D-1$ vanishes and 
the scaling exponent of $|t_{\nu,M}|^2$ is doubled. 
Inserting $M=2$ in Eq.~(\ref{gamma2}) gives $2 \gamma_2(2)=\beta_w$ as 
the exponent characterizing the deviation from transmission $1$. This result 
is expected since the case $M=2$ corresponds to the situation of a 
perfect infinite wire interrupted by a small impurity characterized 
by the scaling exponent $\beta_w \approx 2(K-1)$.   

In the single impurity problem the fRG approximations 
$\beta_s$ for $2\alpha_B=2(1/K-1)$ and $\beta_w$ for $2(K-1)$
are correct to order $U$. To leading order $2(K-1) \approx - 2(1/K-1)$
[see Eq.~(\ref{Kpert})] and the
second term in Eq.~(\ref{gamma2}) is of order $U^2$. 
As terms of order $U^2$ are only partly included in our approximation 
scheme it is questionable if a term similar to the second one will 
be present in the (unknown) exact expression for $\gamma_2$. To derive 
the dependence of $\gamma_1$ and $\gamma_2$ on $K$ and $M$ 
presents a challenge for any method which does not require 
approximations in the strength of the interaction.

\begin{figure}[bt]
\begin{center}
\includegraphics[width=0.4\textwidth,clip]{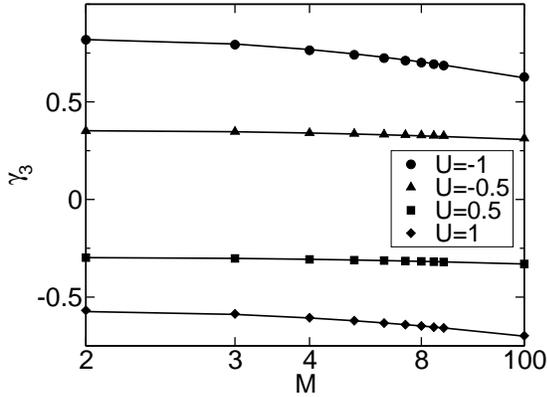}
\end{center}
\caption{Scaling exponent of the conductance 
  for a small on-site energy
  as a function of $M$ (symbols) for different $U$. Note the 
  reciprocal scale of  the $x$-axis. The lines show the fit
  Eq.~(\ref{gamma3}).} 
\label{smallV}
\end{figure}

\subsection{A symmetric junction with a dot site energy}
\label{sub2}

As our second specific case we study a symmetric $M$-leg junction with
$t_\nu=\tilde t$ for $\nu=1,\ldots,M$ and a non-vanishing dot site
energy $V$. Then, due to symmetry, all $\Omega_\nu(0)$ and $\Delta_\nu(0)$
[see Eq.~(\ref{aux})] are equal and we suppress the index $\nu$. 
Eq.~(\ref{taux2}) simplifies to
\begin{eqnarray}
\label{taux2a}
|t|^2  =  \frac{4}{M^2} \; 
\frac{M^2 \Delta^2(0)}{\left[V+M \Omega(0)\right]^2 +M^2  \Delta^2(0)} \; .
\end{eqnarray}
The transmission is determined by the single complex parameter
\begin{eqnarray}
\label{smallgdef} 
g=V+M \Omega(0) - i  M \Delta(0) = - \, 
\frac{1}{{\mathcal G}_d(0+i0)}\; , 
\end{eqnarray}
which itself is a function of
the junction parameters $(\tilde t, V)$ and $M$. 
Via the RG flow of the self-energy for $U \neq 0$ it moreover 
develops a dependence on $\delta_N$ and the interaction $U$. 
The RG flow can nicely be visualized by plotting 
$g$ in the complex plane with $\delta_N$ as a 
parameter.\cite{Enss04,Xavier}  This is done 
in Figs.~\ref{planeflow}a) to \ref{planeflow}c) for $M=2,3,10$,
$V=0.5,1,1.5,2$, and different $\tilde t \in [0.01,0.7]$. 
For decreasing $\delta_N \in [5 \cdot 10^{-4},2.5\cdot 10^{-1}]$ 
and $U \neq 0$ each fixed parameter set $(\tilde t,V)$ leads to 
a flow line. The general form of the 
flow diagrams is independent of the absolute value of $U$. 
The data of the figures were
calculated for $U=-1$, which leads to the flow direction indicated by the
arrows. For $U>0$ the direction is reversed.
On the line $\mbox{Im}\,g=0$ the conductance vanishes and the 
$x$-axis forms a line of ``decoupled chain'' FPs.
For $\mbox{Re}\, g =0$ the transmission is $4/M^2$ and all
points on the $y$-axis are ``perfect junction'' FPs (line of FPs). 
The  ``perfect junction''  transmission $4/M^2$ is also reached 
at $g=\infty$ provided that this point is approached such 
that $-\mbox{Im}\,g  \to \infty$ and $\mbox{Re}\, g $ goes to a 
constant.

For $M=2$ the flow approximately follows a section of a circle 
centered around the origin with a radius $V$. 
For $U<0$ the direction is counter-clockwise and the line 
of ``perfect junction'' FPs ($y$-axis) is stable. The on-site energy 
$V$ does not get renormalized in the RG flow as the interaction 
between the dot site and the first sites of the wires is assumed to 
be $0$. The perfect transmission $4/M^2$ does thus not follow from 
a decrease of $V$ during the RG flow but is a consequence of the
spatial structure of the self-energy, which leads to a 
a resonance at the chemical potential.
For $U>0$ the flow is clockwise and the line of 
``perfect junction'' FPs is unstable.
The system flows to the line of stable 
``decoupled chain'' FPs. The vanishing of the
conductance is a consequence of the long-range oscillatory 
dependence of the self-energy on $j$. 
For $M \geq 3$, $U<0$,  and $\delta_N \to 0$, -$\mbox{Im}\,g$ diverges while 
$\mbox{Re}\, g$ goes to a constant, which implies that the system 
reaches a ``perfect junction'' FP.  
For $U>0$ all trajectories approach the $x$-axis 
at $V$ and the conductance vanishes (line of ``decoupled chain''
FPs). 
To determine scaling exponents we next 
separately consider small and large $V$.

\subsubsection{A small on-site energy}

For small $V$, the transmission between all the equivalent
wires is close to the perfect value $4/M^2$. The dependence on
$\delta_N$ can be described by a power-law. In Fig.~\ref{smallV} the
scaling exponent as a function of $M$ is shown for different $U$. 
To leading order in $U$ it is independent of $M$. 
The exponent can be fitted by 
\begin{eqnarray}
\label{gamma3}
\gamma_3(M)=-\beta_s + 4 \frac{\beta_w+\beta_s}{M} 
\left(1-\frac{1}{M}\right) \; .
\end{eqnarray} 
In accordance with the stability properties of the line of 
``perfect junction'' FPs discussed in connection with 
Figs.~\ref{planeflow}a) to \ref{planeflow}c), $\gamma_3 <0$ for $U>0$
and $\gamma_3 > 0$ for $U<0$.
The $M=2$ case
is equivalent to the problem of a single weak site impurity
interrupting an otherwise perfect infinite chain 
and $\gamma_3(2)$ is equal to the
respective scaling exponent $\beta_w$ (similar as for a slightly
modified link in Sect.~\ref{sub1}). 
Within our approximation scheme the second term in Eq.~(\ref{gamma3})
is thus important to reproduce a result obtained earlier for $M=2$. 
As it is of order $U^2$ it is nonetheless unclear if a similar term 
occurs in the (unknown) exact expression for $\gamma_3$.

\subsubsection{A large on-site energy} 

In the limit of large $V$ the transmission between all the wires is
small and we analyze the scaling with respect to zero transmission. 
The scaling exponent is independent of $M$ and up to our numerical 
accuracy given by 
\begin{eqnarray}
\label{gamma4}
\gamma_4=\beta_s \; ,
\end{eqnarray}
i.e.~the strong impurity (weak link) exponent for
a single infinite wire (see Fig.~5 of Ref.~\onlinecite{Enss04}). 

\subsection{Fixed points and renormalization group flow for general
  junction parameters}
\label{sub3}

The RG flow, FP structure, and scaling exponents for arbitrary
junction parameters can be understood based on the results obtained 
for the above two classes of junction parameters. To shorten 
the discussion we here focus on the more important case $U>0$. Results for
$U<0$ can be deduced by inverting the direction of the flow. 

As found in subsections  \ref{sub1} and \ref{sub2} the 
``perfect junction'' FP of the $M$-leg system with effective 
transmission $4/M^2$ is unstable towards the two possible 
perturbations in which only one 
junction parameter is modified: (i) a single modified hopping and (ii) 
a non-vanishing on-site energy. These instabilities are characterized by
the two exponents $\gamma_2(M)$ (modified link) and 
$\gamma_3(M)$ (on-site energy). 

In the case (i), for $M \geq 3$, and $\tilde t$ larger than the 
single modified hopping the system flows to the  ``perfect junction'' 
FP for $M-1$ legs. For $\delta_N \to 0$ the conductance across the
modified link vanishes with exponent $\gamma_1(M)$, while 
the conductance within the subsystem of the $M-1$ equivalent wires 
approaches $4/(M-1)^2$ with the same exponent.
For $M=2$ the flow is to the ``decoupled chain'' FP  and the
conductance vanishes with 
exponent $\gamma_1(2)=\gamma_4=\beta_s$. 
If the hopping between the first site of wire $\nu = M$ 
and the dot site is larger than $\tilde t$ for 
$\delta_N \to 0$ and all $M \geq 2$ the ``decoupled chain'' FP 
is reached. 
To analyze the scaling of the conductance close to this 
FP we start out from
Eq.~(\ref{taux2}) and use an expansion similar to 
Eqs.~(\ref{e:low_link1}) and (\ref{e:low_link2}). The small parameter
of the expansion is $D=\Delta/\Delta_M$. The numerics shows that it 
asymptotically vanishes as $\delta_N^{\beta_s}$. 
For $\nu \leq M-1$ this leads to 
\begin{eqnarray}
\label{tlarger1}
G_{\nu,M} \propto  \delta_N^{\gamma_4}
\end{eqnarray}
while for $\nu,\nu' \in [1,M-1]$ we find
\begin{eqnarray}
\label{tlarger2}
G_{\nu,\nu'} \propto  \delta_N^{2 \gamma_4} \; .
\end{eqnarray}
In the case (ii) the RG flow also 
ends at the ``decoupled chain'' FP and all conductances vanish with 
exponent $\gamma_4$.

We next perturb the ``perfect junction'' FP of the $M$-leg system 
by slightly modifying more than one of the hopping matrix
elements. If $2 \leq M_1 < M-1$ of the $t_\nu$ are reduced compared 
to the hopping $\tilde t$ across the remaining $M_2=M-M_1$ links 
the system flows to the ``perfect junction'' FP of the $M_2$-leg 
system. If $M_1$ of the $t_\nu$  are increased compared to $\tilde t$ the 
system generically approaches the  ``decoupled chain'' FP. 
This behavior is changed if $\tilde M_1 \geq 2$ of the $M_1$ 
increased hoppings are equal and larger than all the others. In 
this case the system flows to the  ``perfect junction'' FP for 
$\tilde M_1$ legs. If all $t_{\nu}$ are different the ``decoupled
chain'' FP is reached.

If the perturbation of the ``perfect junction'' FP of the $M$-leg 
system consists of an on-site energy $V$ and one or more 
modified hoppings the system flows to the ``decoupled chain'' FP. 

From these considerations we conclude that for the majority of
possible junction parameters of the $M$-wire junction, 
for $\delta_N \to 0$ the system flows to the ``decoupled chain'' FP. 
On asymptotically small scales the conductance vanishes as
\begin{eqnarray}
\label{generic}
G_{\nu,\nu'} \propto \delta_N^{\xi \beta_s} \; . 
\end{eqnarray}
Depending on the junction parameters as well as on the wire indices
$\nu$ and $\nu'$, $\xi$ might be $1$ or $2$ (for examples, see above).

Only for $V=0$ and if $M_{\rm max} \geq 2$ of the $M$ links between the 
dot site and the first lattice sites of the wires have hopping $t_{\rm
  max}$, where $t_{\rm max}=\max_{\nu \in [1,M]}\{t_{\nu}\}$, the
system flows to the ``perfect junction'' FP for $M_{\rm max}$ legs. 
In this case the power-law scaling of the conductance between two wires 
coupled to the dot with hopping $t_{\rm max}$ is given by
\begin{eqnarray}
\label{special1}
\frac{e^2}{h} \, \frac{4}{M_{\rm max}^2} 
- G_{\nu,\nu'} \propto \delta_N^{\beta_s/M_{\rm max}} \; .  
\end{eqnarray}
The conductance between one wire coupled by $t_{\rm max}$ and the 
other by $t_{\nu} < t_{\rm max}$ vanishes as 
\begin{eqnarray}
\label{special2} 
G_{\nu,\nu'} \propto \delta_N^{\beta_s/M_{\rm max}} \; ,  
\end{eqnarray}
while it goes like 
\begin{eqnarray}
\label{special3} 
G_{\nu,\nu'} \propto \delta_N^{2 \beta_s/M_{\rm max}} \; .  
\end{eqnarray}
if both wires $\nu$ and $\nu'$ have a hopping  to the dot site 
smaller than $t_{\rm max}$. As in Eq.~(\ref{tlarger2}) the factor $2$ in 
Eq.~(\ref{special3}) follows from an expansion similar to the 
one used in Eqs.~(\ref{e:low_link1}) and (\ref{e:low_link2}).

\section{Summary and Perspectives}
\label{summary}

In this paper we studied the conductance $G_{\nu,\nu'}$ for a dot
junction of $M$ semi-infinite quantum wires. The junction as well as 
the wires are described by a microscopic lattice model. In a finite
section of $N$ sites the fermions in each wire are modeled to
interact via a nearest-neighbor interaction which smoothly vanishes 
close to the lattice sites 
$j=N$ (smooth contact). Investigating the scaling of the
conductance as a function of $\delta_N \propto 1/N$ 
led to a comprehensive picture 
of the FP structure, scaling exponents, and RG flow of the model
studied. We mainly used an approximation scheme that is based on the
fRG method and provides reliable results for week to intermediate
interactions with $1/2 \leq K \leq 3/2$. At half-filling this  
corresponds to bulk nearest-neighbor interactions $-1.5 \leq U \leq 2$.  
Additional insights were obtained using the HFA. 

Compared to the well studied single impurity problem for 
$M \geq 3$ the low-energy physics of the $M$-wire junction is much 
richer, allowing for a variety of FPs and scaling exponents. 
Furthermore for $M \geq 3$ one of them, the ``perfect junction'' FP,
is characterized by two exponents: $\gamma_2(M)$ and $\gamma_3(M)$.
They can individually be read off from the scaling of the conductance 
if the FP is perturbed in a specific way.     
In contrast to the single impurity problem, in which the FP reached is
solely determined by the sign of the interaction, for
junctions of three and more wires this in addition depends 
on the junction parameters. Depending on
the wire indices between which the conductance is calculated the
scaling exponents close to a FP might differ by factors of two. This
can have two reasons: (i) In an expansion of the effective
transmission in terms of a small parameter that carries the power-law
scaling [see e.g.~Eqs.~(\ref{e:low_link1}) and (\ref{e:low_link2})],
depending on $\nu$ and $\nu'$ the first or second order term might 
be the first non-vanishing one. (ii) For certain $M$ the prefactor of the
leading order term might vanish. 

Junctions and networks of TLLs were earlier 
studied.\cite{Nayak,Safi,Lal,Chen,Claudio,Das,Doucot} 
The method and model used in Ref.~\onlinecite{Lal} come closest to
ours. The authors apply a fermionic poor-man's like RG originally developed 
for the single impurity problem to the three- and four-leg 
junction.\cite{MatveevGlazman} They consider a microscopic dot
junction model to motivate the investigation but then leave the
framework of this model when studying the RG flow of an
effective $S$-matrix. The results obtained in this
paper for three legs are partly equivalent to ours. 
In Refs.~\onlinecite{Enss04} and
\onlinecite{VMresotun} a detailed account of the differences of the  
poor-mans RG and our method is given for the single impurity case and 
resonant tunneling in a TLL. 

Using the fRG and HFA we obtained expressions for
the scaling exponents which we expect to be correct (at least) to
leading order in $U$. It is very desirable to derive  
the exact $K$- and $M$-dependence of these exponents.
For increasing complexity of the junctions obtaining such
expressions for simplified, effective, but still generic models 
requires very sophisticated methods, even if only specific parts of the
parameter space are considered.\cite{Claudio} 

We considered a dot junction model in which the interaction
between fermions on the dot site and the first lattice sites of the
wires is set to zero. We also investigated the case in which this 
interaction does not vanish. The additional nearest-neighbor
interaction across the $M$ bonds does not alter the results presented
here.

An extension of our method to the case in which the strength
of the bulk interaction depends on the wire index $\nu$ is
straightforward and might lead to new insights. As 
exemplified in Ref.~\onlinecite{Xavier} the fRG can also be used to
investigate other types of junctions (e.g.~ring like
geometries), with different FPs and new scaling exponents. 

For the complex junction studied here the temperature $T$
and the infrared cutoff $\delta_N$ present equivalent scaling
variables only on asymptotically small scales.
The fRG method can be set up for finite $T$ and leads to reliable 
results also for intermediate to large temperatures.\cite{Enss04,VMresotun} 
For two wires  our model is equivalent to the one considered to study 
resonant tunneling through a quantum dot embedded in a TLL, with a one 
lattice site dot. In this case, for fixed $N$, and 
$\delta_N \leq T \leq B$ the conductance as a function of $T$ shows 
a very rich behavior. Depending on the dot parameters, temperature
regimes in which $G(T)$ follows ``universal'' power-laws 
as well as non-universal regimes were identified.\cite{Enss04,VMresotun}
We thus expect to find similar rich behavior for $M \geq 3$.
The conductance as a function of temperature is
easily accessible in transport experiments. Investigating
$G_{\nu,\nu'}(T)$ might thus also become important for the 
interpretation of future transport experiments on junctions of 
quasi one-dimensional quantum wires. We will present results for 
$G_{\nu,\nu'}(T)$ in an upcoming publication.

\section*{Acknowledgments}

We thank S.~Andergassen, T.~Enss, and W.~Metzner for valuable 
discussions. X.B.-T. was supported by a Lichtenberg-Scholarship of the 
G\"ottingen Graduate School of Physics. V.M. and K.S. are grateful
to the Deutsche Forschungsgemeinschaft (SFB 602) for financial
support.

\end{document}